# Loss Allocation in Submarine Armored Three-core HVAC Power Cables

Juan Carlos del-Pino-López  
Dpto. Ingeniería Eléctrica  
Escuela Politécnica Superior  
Virgen de África 7  
41011, Sevilla, Spain  
vaisat@us.es  

Pedro Cruz-Romero  
Dpto. Ingeniería Eléctrica  
Escuela Técnica Superior de Ingeniería  
Avda. de los Descubirmientos s/n  
41092, Sevilla, Spain  
plcruz@us.es  

Luis Carlos Sánchez-Díaz  
Dpto. Ingeniería Eléctrica  
Escuela Politécnica Superior  
Virgen de África 7  
41011, Sevilla, Spain  
lsanchezd@us.es

*Abstract—* Loss allocation of the three different components (conductor, sheaths and armor) of solidly bonded three-core separated lead-sheathed armored cables, frequently employed in offshore wind farms, is challenging due to the lack of accurate enough analytical expressions in the IEC standard. Also, loss allocation through experimental tests leads to inaccurate results since it is based on questionable assumptions. This paper improves both the IEC formulae and experimental methods by means of different analytical corrections in the conductor and sheath loss expressions. To this aim, an ad hoc application interface (Virtual Lab) based on 3D numerical simulations (finite element method) has been developed. This tool virtualizes and automates different test setups to emulate, in few seconds, the most employed experimental procedures in this type of cable. The analytical corrections have been derived from an in-depth analysis of a first set of 368 cables, ranging from 30 to 275 kV. The new loss expressions were successfully applied to a second set of 645 armored cables of quite diverse features (voltages from 10 to 275 kV, sections and dimensional parameters), hence bringing a general framework for any kind of three-core armored cable.

## I. INTRODUCTION

One of the major requirements in the project of a power cable is to avoid overheating it during its operating life. This fact is even more critical in the case of submarine cables due to their cost and difficulties to repair [1]-[7]. This topic was dealt with by the authors in a previous paper [1], largely improved in this paper with additional findings and results, as outlined throughout the text.

In this kind of cable three metallic parts are the sources of power losses [1]-[6]: conductors ($P_c$), metallic sheaths ($P_s$) and armor ($P_a$) (Fig. 1), so a correct thermal rating estimation requires a proper allocation of losses in these three heat sources. To this purpose a lumped-parameter thermal model is customarily used [2]-[4], where the three loss components are employed as inputs. This is the approach of the IEC (International Electrotechnical Commission) standard [2], where the allocation of power losses is carried out by means of analytic expressions of loss factors $\lambda_1$ and $\lambda_2$:

$$\lambda_1 = \frac{P_s}{P_c}, \qquad \lambda_2 = \frac{P_a}{P_c}. \qquad (1)$$

The induced currents in sheaths and armor can just close themselves as eddy currents inside each of the three sheaths and the armor (bonded at a single point SP) or also circulate outside each sheath/armor (bonded at both ends, solid bonding SB). In [2], $\lambda_1$ is expressed as the sum of circulating-currents ($\lambda'_1$) and eddy-currents ($\lambda''_1$) loss subfactors. However, and focusing on wire-armored separated-lead sheathed three-core cables, [2] assumes that $\lambda''_1$ is neglectable when sheaths are in SB (preferred configuration in submarine applications [6]). Having this in mind, [2] provides the following expressions for $\lambda_1$ and $\lambda_2$:

$$\lambda_1 = \lambda'_1 = \frac{R_s}{R_c} \frac{1.5}{1 + \left(\frac{R_s}{X}\right)^2}; \quad X = 2\omega 10^{-7} \ln \frac{2s}{d}, \qquad (2)$$

$$\lambda_2 = 1.23 \frac{R_a}{R_c} \left(\frac{2c}{D_a}\right)^2 \frac{1 - \frac{R_c}{R_s} \lambda'_1}{\left(\frac{2.77 R_a 10^6}{\omega}\right)^2 + 1}, \qquad (3)$$

being $R_s$ and $X$ the AC resistance and reactance of each sheath (Ω/m), respectively, $R_c$ the AC resistance of each conductor (Ω/m), $s$ the distance between conductor axes (mm), $d$ the mean diameter of each sheath (mm), $R_a$ the armor resistance (Ω/m), $c$ the distance between the center of a conductor and the cable axis (mm), $D_a$ the mean diameter of armor (mm) and $\omega$ the angular frequency of the system.

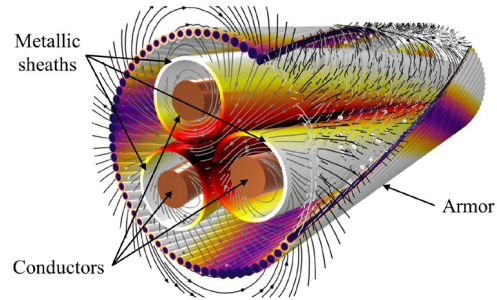

Fig.1. Main components of an armored lead-sheathed three-core cable, power losses distribution and magnetic flux lines.

Nonetheless, these expressions ignore the influence of the relative twisting of power cores and armor wires (Fig. 1), giving rise to calculated armor losses higher than expected from the field experience [8]-[11]. In this sense, to better characterize $P_a$, two experimental procedures have been introduced [12], [13], both based on experimentally measuring the total power dissipated by the SB cable either or not with armor. Given that it is not possible to measure the power losses in each cable element separately, both procedures obtain the sheath and conductor losses by [2], following several simplifying assumptions (to be described in the next section), what entails inaccurate results in the experimental evaluation of $P_a$. The underlaying reason for these results is found by using numerical simulations based

This work was supported by research grants ENE2017-89669-R (AEI/FEDER, UE) and 2018/00000740 (VI PPIT-US).

on the finite element method (FEM) (either in 2D [8]-[10], [14]-[16] or 3D [17]-[23]), that show how [2] introduces errors in the estimation of the sheath losses [1], [10], [14]-[17], since it is assumed that eddy currents are negligible in SB cables. Furthermore, these studies also underline the noticeable influence of the armor, and especially its magnetic properties, on the conductor losses [1], [21], [24], something missing in [2].

Therefore, it is clear the need of improving the IEC estimation of sheath and conductor losses by [2], that would also enhance the results of the experimental method (EM) of [12], [13]. This is tackled in this paper in an integrative way by using advanced 3D-FEM simulations. With this aim, a user-friendly application interface, named Virtual Lab (VL) was developed in [1]. This tool (largely improved here) exploits the benefits of periodic geometries in 3D-FEM simulations [20] for emulating the laboratory experimental tests, so that virtual measurements in different setups (cables with and without armor and different bonding schemes) can be obtained. Through this application, an increase in the loss determination accuracy over previous EM [12], [13] was reached in [1] by the introduction of two numerical correction factors in the analytical expressions of the conductor and eddy-current losses provided by [2]. However, these factors depend on geometrical and electrical parameters of the cable, such as the armor magnetic permeability and the lay length of conductors and armor wires, so they are suitable only for a small set of three-core armored cables.

In this work, thanks to new improvements in the VL (able to run virtual tests in few seconds) these numerical correction factors are now replaced by two analytical expressions derived from the analysis of up to 368 three-core armored and unarmored cable configurations, providing a general framework (valid for the great majority of the cables) to compensate the errors in the evaluation of the conductor and sheath losses, hence bringing an important improvement in the experimental evaluation of armor losses in laboratory setups. Furthermore, the new features of the VL are not only intended to help in preparing experimental setups, but also in analyzing and optimizing cable design.

The paper is structured as follows. In section II the EM is described in more detail. Section III briefly describes the improvements in the 3D-FEM model and the VL application developed to automate all the analyses. In Section IV important conclusions regarding the power losses and the IEC standard are derived from the VL by analyzing a set of 368 HVAC three-core armored and unarmored cables. These results are later considered in section V for proposing a general framework that improves the accuracy in the EM-based armor losses estimation, showing excellent results in an additional set of 645 three-core armored cables in SB. Eventually, in section VI some concluding remarks are included.

## II. EM FOR ALLOCATING POWER LOSSES

For the experimental evaluation of the armor losses in three-core submarine cables, previous studies [12], [13] proposed a difference measurement method to minimize the influence of errors in measurements, hence improving the accuracy of the estimated values. It is based on testing two identical pieces of the same cable, but one unarmored (subscript 0) and the other armored (subscript 1). The power balance equations involved in both tests are

$$P_{m0} = P_{c_0}^J + P_{s_0}^J + P_{s_0}^{ec}, \qquad (4)$$

$$P_{m1} = P_{c_1}^J + P_{s_1}^J + P_{s_1}^{ec} + P_a, \qquad (5)$$

being $P_m$ the measured total power, $P^J$ the Joule losses due to circulating currents, $P^{ec}$ the Joule losses due to eddy currents and $P_a$ the total losses in the armor wires (caused by eddy currents and hysteresis as verified in [8-11]). Also, other measurements are taken during tests, such as the DC resistance of each conductor ($R_c^{DC}$), each sheath ($R_s^{DC}$) and the armor ($R_a^{DC}$), as well as the conductor ($I_c$) and sheath currents ($I_s$), assuming that sheaths and armor are solidly bonded. In [13] all measurements are performed at ambient temperature ($\theta_{amb}$) and in a very short period to prevent the impact of thermal heating on losses. Then, for obtaining the armor losses $P_a$, a difference method is applied by subtracting (4) from (5):

$$P_a = \Delta P_m - \Delta P_c^J - \Delta P_s^J - \Delta P_s^{ec}, \qquad (6)$$

with

$$\Delta P_m = P_{m1} - P_{m0}, \qquad (7)$$

$$\Delta P_c^J = P_{c_1}^J - P_{c_0}^J, \qquad (8)$$

$$\Delta P_s^J = P_{s_1}^J - P_{s_0}^J, \qquad (9)$$

$$\Delta P_s^{ec} = P_{s_1}^{ec} - P_{s_0}^{ec}. \qquad (10)$$

However, most terms in (7)-(10) cannot be evaluated separately through an experimental setup, so they are obtained by considering the following assumptions:

- For both the armored and unarmored cables, the AC resistance of the conductors is derived from the measured DC resistance by including the skin ($y_s$) and proximity effects ($y_p$) factors as defined in [2]. This way, the total losses in the conductors are obtained using the measured current as

$$P_c^J = 3R_c^{DC}(1 + y_s + y_p)I_c^2. \qquad (11)$$

- The AC resistance of the sheaths is assumed to be equal to the measured DC resistance. Thus, the Joule losses due to circulating currents in the sheaths are derived from the measured sheath current as

$$P_s^J = 3R_s^{DC}I_s^2. \qquad (12)$$

- It is assumed that sheath eddy-current losses in three-core cables (either armored or unarmored) are comparable to those obtained in the case of three single-core unarmored cables in trefoil formation (there are no formulae available in [2] for the case of three-core cables). Thus, the corresponding loss factor is employed:

$$\lambda_1'' = \frac{R_s^{DC}}{R_c}\left[g_s\lambda_0(1 + \Delta_1 + \Delta_2) + \frac{(\beta_1 t_s)^4}{12 \cdot 10^{12}}\right] \qquad (13)$$

where $t_s$ is the sheath thickness, and $g_s$, $\lambda_0$, $\beta_1$, $\Delta_1$ and $\Delta_2$ are parameters defined in [2].

These assumptions involve important inaccuracies. For example, if the same current is injected through the conductors during both tests ($I_{c0} = I_{c1} = I_c$), then the first hypothesis implies that $\Delta P_c^J = 0$, since it is considered that $R_c$ is the same for the armored and unarmored cables. However, in [1], [8]-[11], [21], [24] it is observed that $R_c$ is affected by the presence of the armor, and consequently $\Delta P_c^J \neq 0$. On the other hand, another controversial point is regarding the last assumption, since there is no evidence to support that hypothesis, and in practice it leads to assume that $P_{s_1}^{ec} = P_{s_0}^{ec}$ in the tested cables, hence having $\Delta P_s^{ec} = 0$. Nonetheless, [10], [14]-[17] showed that the presence of the armor influences the current density distribution in the sheaths, so that probably $\Delta P_s^{ec} \neq 0$. As a result, the estimation of $P_a$ derived from (6) is inaccurate since it contains part of the losses associated to the conductors and the sheaths, pointing out the need of improving the estimation of conductors and sheaths losses for obtaining accurate results of $P_a$. To this aim, 3D-FEM simulations are employed, as detailed next.

### III. 3D-FEM MODEL: VIRTUAL LAB

Thanks to recent advances, the use of 3D-FEM simulations for analyzing complex three-core armored cables is now more feasible [1], [17]-[23], not being required the use of clusters or workstations plenty of RAM memory to obtain a solution. In particular, [20] proposed the use of a shorter periodic geometry for the three-core armored cable, reducing the computational requirements significantly, being possible to run the simulations in computers with less than 64 GB of RAM memory in about 30 minutes, and all this providing accurate results when compared to experimental measurements. This improvement was possible since all the electromagnetic interactions involved in three-core armored cables are present in a cable length ($L$) as short as the so called "crossing pitch" ($CP$):

$$CP = \frac{1}{\frac{1}{L_a} + \frac{1}{L_c}}, \quad (14)$$

valid for cables where the armor and the phases are twisted in opposite directions (contralay), being $L_c$ and $L_a$ the lay length of phases and armor wires, respectively, and $CP$ the distance in which an armor wire makes a full revolution around a particular phase. Nonetheless, a subsequent in-depth analysis in [1,22,23] concluded that the length of the model can be further reduced up to

$$L = \frac{CP}{N}, \quad (15)$$

where $N$ is the number of armor wires. This fact reduces the geometry to be modelled to just a small slice of the cable (Fig. 2), so the simulation time that is now below 1 minute in computers with less than 16 GB of RAM memory. This is achieved by applying appropriate periodic boundary conditions at both ends of the 3D model, where the relative rotation ($\theta$) between the coordinate systems assigned to the source ($\vec{e}_x^s, \vec{e}_y^s$) and destination boundaries ($\vec{e}_x^d, \vec{e}_y^d$) must be considered, defined as (Fig. 2)

$$\theta = \frac{2\pi L}{L_c}. \quad (16)$$

Having all this in mind, the electromagnetic problem is analyzed by solving the following equation:

$$\nabla \times \left(\frac{1}{\mu} \nabla \times \vec{A}\right) + j\omega\sigma\vec{A} = \vec{J}_e, \quad (17)$$

where $\vec{A}$ is the magnetic vector potential, $\sigma$ the conductivity (temperature dependent), $\mu$ the magnetic permeability and $\vec{J}_e$ the external current density. It should be noticed that a complex magnetic relative permeability $\mu_r$ is considered for the armor wires to take into account hysteresis losses [25]:

$$\mu_r = \mu' - j\mu''. \quad (18)$$

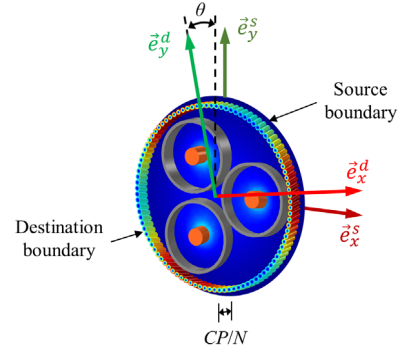

Fig. 2. Shortened 3D-FEM geometry and rotation between source and destination boundaries for implementing periodic boundary condition.

Once solved, the volumetric Joule losses generated in the conductors, sheaths and armor can be derived from the current density $\vec{J}$ as

$$P_J = \int \frac{\vec{J} \cdot \vec{J}^*}{\sigma} d\Omega. \quad (19)$$

Also, magnetic losses in the armor ($P_{mag}$) are obtained by

$$P_{mag} = \omega\mu_0\mu'' \int \vec{H} \cdot \vec{H}^* \, d\Omega, \quad (20)$$

where $\mu_0$ is the magnetic permeability of vacuum, and $\vec{H}$ the magnetic field.

An example of the improvement achieved by the proposed approach is shown in Table I, where the most relevant data and the main results obtained in the case of a 115 kV, 240 mm² three-core armored cable are compared for a model length equal to $CP$ and $CP/N$, where $R_p$ and $X_p$ are the positive-sequence resistance and inductive reactance of the cable, respectively. Also, relevant data regarding mesh size and the RAM memory employed for the simulations are included. Simulations have been performed in a laptop with an i7 processor and 64 GB of RAM memory using COMSOL Multiphysics® [26].

TABLE I. RESULTS IN A 115 KV, 240 MM² THREE-CORE ARMORED CABLE

| Parameter | $L = CP$ | $L = CP/N$ |
|---|---|---|
| Simulation time (s) | 620 | 30 |
| Mesh elements | 887260 | 94442 |
| RAM (GB) | 60 | 4 |
| $R_p$ (Ω/km) | 0.0860 | 0.0858 |
| $X_p$ (Ω/km) | 0.1585 | 0.1592 |
| $I_s$ (A) | 106.25 | 107.4 |
| $P_c$ (W/m) | 50.265 | 50.094 |
| $P_s$ (W/m) | 12.208 | 12.002 |
| $P_a$ (W/m) | 2.0857 | 2.0773 |

As can be seen, a reduction in the computation time of about 95 % is obtained with negligible effects on the accuracy, with just 30 s for solving such complex model and with negligible effects on the accuracy. Thus, through this new improvement in the electromagnetic simulation of three-core armored cables, the VL presented in [1] is now updated, as described next.

*A. Virtual Lab*

In [1] a stand-alone and user-friendly application was presented for helping in the analysis and design of three-core armored cables. This tool, based on the COMSOL Multiphysics®-integrated Application Builder, provides in advance valuable data for improving the preparation and, hence, the accuracy of lab tests. Now, the VL has been updated with new and improved features to emulate, not only the steps taken in the EM presented earlier, but also typical experimental setups usually employed for characterizing this type of power cables [8]-[13]. The most relevant are:

- Cable geometry can be updated manually or by uploading a text file with all the data. Additionally, a set of 20 cables ranging from 30 kV to 275 kV are available as templates (Fig. 3a). Once done, the geometry and meshing are automatically updated.
- A list of materials is available for conductors (aluminum or copper), sheaths (lead) and armor wires (galvanized or stainless steel), including real or complex permeability.
- Additional data may be included to emulate the laboratory settings, such as the ambient temperature, the maximum current of the power source, the length of the cable to be tested, as well as the resistance of the load resistor employed during tests (Figs. 3a-3c).
- A DC analysis can be performed in the cable to obtain $R_c^{DC}$, $R_s^{DC}$ and $R_a^{DC}$. Important data relative to the voltage, current and power dissipated during tests are also provided. Graphical schemes of the performed test are also included (Fig. 3b).
- Up to 4 AC tests can be performed in the cable (with and without armor, considering sheaths and armor in SP or SB) providing "virtual measurements" (VM) of those usually obtained in real setups (total power losses, applied voltage, sheath and conductor currents, and induced voltages in sheaths (only in SP)). Graphical schemes of the performed test are also included (Fig. 3c).
- Additional data derived from VM are provided for each AC analysis, such as the cable resistance and reactance (Fig. 3c).
- The application also evaluates the losses in conductors, sheaths and armor separately. The distribution of these losses within all metallic parts of the cable are plotted (3D-color plot of Fig. 3c). Additional plots can be also represented for evaluating any other magnitude.
- In all the AC studies the loss factors derived from the results and the IEC standard are compared, showing the relative differences (Fig. 3c).
- All the studies can run at once, so that a summary of all direct and indirect virtual measurements, as well as estimated parameters are presented at the end.
- AC studies can be performed at different frequencies to analyze the harmonics influence on the electrical parameters of the cable.
- Parametric sweeps can be now performed faster to analyze the influence of a set of geometrical and/or material parameters on the cable behavior.
- Finally, it includes a specific section where the methodology presented in Section II to estimate the armor power losses is replicated based on the VM derived from the virtual setups, summarizing the differences relative to the 3D-FEM results.

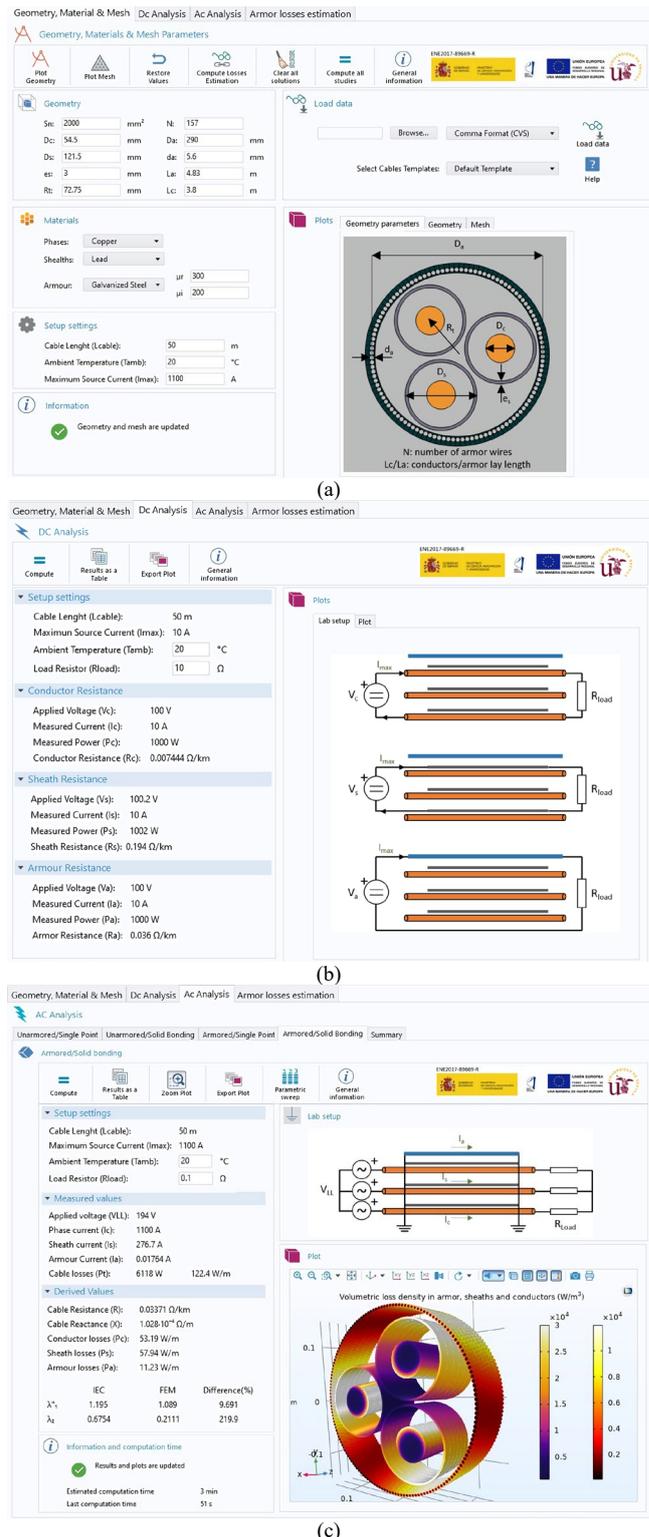

Fig. 3. Virtual lab sections: a) Cable data and geometry, b) DC analysis and c) AC analysis with results for armored cable in solid bonding.

IV. 3D FEM-BASED VS IEC-BASED LOSSES RESULTS

Since the EM in [13] is based on the numerical difference of $P_c$ and $P_s$ for the armored and unarmored cables ((6)-(10)), it is of interest to analyze both arrangements for evaluating how comparable are the values of $P_c$ and $P_s$ derived by [2] and FEM simulations, as a diagnostic of the inaccuracies of the EM. By means of the VL, important conclusions are derived from the analysis of a set of 23 geometries of HVAC real cables, ranging from 30 kV to 275 kV. Table II summarizes the main characteristics of some of the studied cables (obtained from [8]-[21]), where $d_c$, $d_s$ an $d_a$ denote the external diameter of the conductor, sheath and armor wire, respectively, $t_s$ is the sheath thickness, and the rest of the parameters were previously defined. All the cases have been simulated assuming all metallic parts at 20 ºC, considering copper and aluminum conductors, with and without armor, as well as sheaths in SP and SB (material properties from [2]). Moreover, three values of $\mu_r$ have been considered (100−50j, 300−200j and 600−350j), resulting in a set of 368 simulated cases.

TABLE II. SUMMARY OF THE MAIN PROPERTIES OF SOME OF THE THREE-CORE ARMORED CABLES ANALYZED

| Volt. (kV) | $I_c$ (A) | $d_c$ (mm) | $d_s$ (mm) | $t_s$ (mm) | $c$ (mm) | $d_a$ (mm) | $D_a$ (mm) | $N$ | $L_c$ (m) | $L_a$ (m) |
|---|---|---|---|---|---|---|---|---|---|---|
| 30 | 200 | 13.4 | 37 | 1.7 | 23.67 | 4 | 97.17 | 69 | 1.4 | 0.9 |
| 115 | 530 | 23.5 | 78.7 | 3.3 | 49.48 | 6 | 196.2 | 98 | 1.5 | 3.1 |
| 132 | 900 | 34.5 | 82.5 | 2.5 | 50.23 | 5.6 | 110 | 204 | 2.6 | 3.4 |
| 150 | 650 | 30.25 | 80.6 | 2.8 | 49.42 | 6 | 195 | 95 | 2.6 | 1.8 |
| 220 | 975 | 49 | 104 | 3 | 62.35 | 5.6 | 250 | 135 | 3.3 | 4.1 |
| 275 | 1100 | 54.5 | 121.5 | 3 | 72.75 | 5.6 | 290 | 157 | 3.8 | 4.8 |

### A. Unarmored Cables

For the unarmored SP-bonding case Fig. 4 shows that [2] generally overestimates the values of $\lambda''_1$ over FEM simulations (differences mostly below 15 %) observing higher differences for the largest cables (in the figure the cable numbering is ordered with increasing voltages and cross sections).

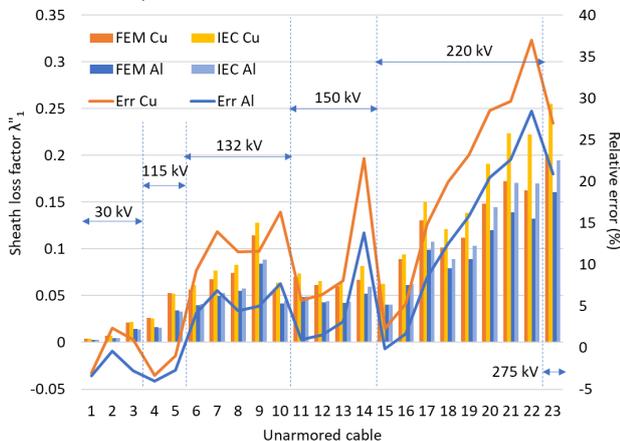

Fig. 4. Unarmored cables in SP: results for $\lambda''_1$ derived from FEM and [2] and relative error ($\theta_{amb} = 20$ ºC).

On the contrary, when sheaths are in SB different results are obtained depending on how $\lambda''_1$ is considered. In this sense, $\lambda''_1$ is usually neglected in SB unarmored cables, as suggested in [2], since circulating losses are expected to be predominant. This assumption leads to the results of Fig. 5, where the sheath losses derived from [2] (assuming $\lambda_1 = \lambda'_1$) and the VL are depicted. It is observed how the analytical approach underestimates the sheath losses, with differences generally greater than 10 %. This is in good agreement with [1], [15], where it is suggested that $P_s^{ec}$ may be proportional to circulating losses, especially in larger cables, so $\lambda''_1$ should not be neglected. If so, better results are obtained (Fig. 6). In this situation, [2] overestimates the sheath losses, mainly below 10 %, even though greater errors are observed precisely in the largest cables.

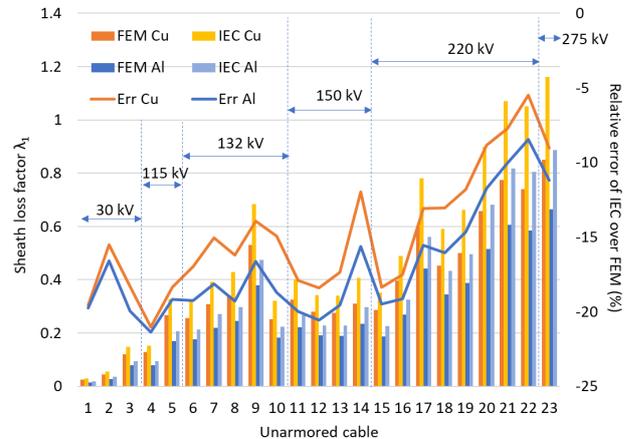

Fig. 5. Unarmored cables in SB: results for $\lambda_1$ derived from FEM and [2] and relative error when $\lambda''_1 = 0$ ($\theta_{amb} = 20$ ºC).

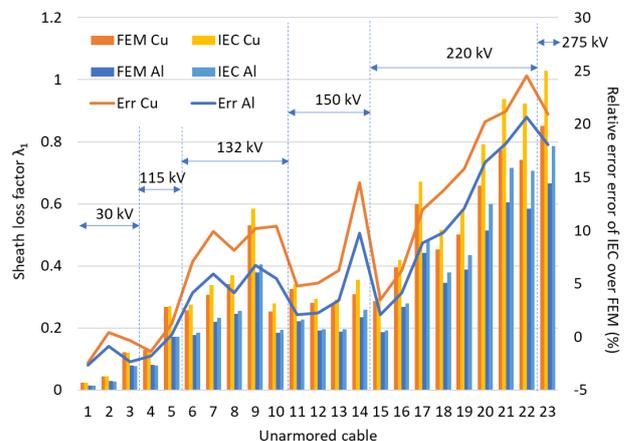

Fig. 6. Unarmored cables in SB: results for $\lambda_1$ derived from FEM and [2] and relative error when $\lambda''_1 \neq 0$ ($\theta_{amb} = 20$ ºC).

Regarding the estimation of $P_c$ in unarmored cables, Fig. 7 shows a good agreement between FEM and [2] results, since relative differences are always below 1 % for all the unarmored cables analyzed, either in SB or SP. Nonetheless, these results suggest that $R_c$ is slightly influenced by the sheath bonding, since higher values in $P_c$ are observed when considering sheaths in SP.

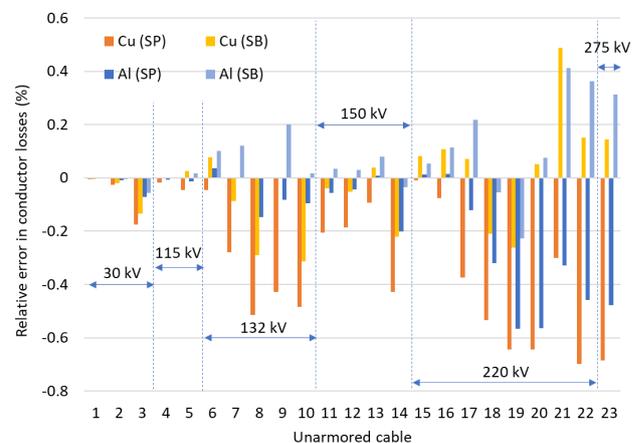

Fig. 7. Unarmored cables in SP and SB: relative error in the computation of $P_c$ from FEM and [2] ($\theta_{amb} = 20$ ºC).

## B. Armored Cables

Interesting results are also derived in the case of armored cables. For example, since there is no analytical expression for the sheath losses when they are in SP, [13] assumes that they are comparable to those derived from [2] for the case of three single-core conductors in trefoil formation. The relative error derived from this approach is depicted in Fig. 8a when considering copper conductors and different values of $\mu_r$ for the armor wires (similar results are observed for aluminum). It is easily seen that this assumption leads to underestimated values of $\lambda_1$ and $P_s$ when compared to FEM results (labeled as SP in Fig. 8a and 8b). Furthermore, in Fig. 8b it is also observed that the sheath losses derived by FEM increase with $\mu_r$ in all the cables, being between 1.3 and 2.5 times higher in armored cables than in unarmored ones.

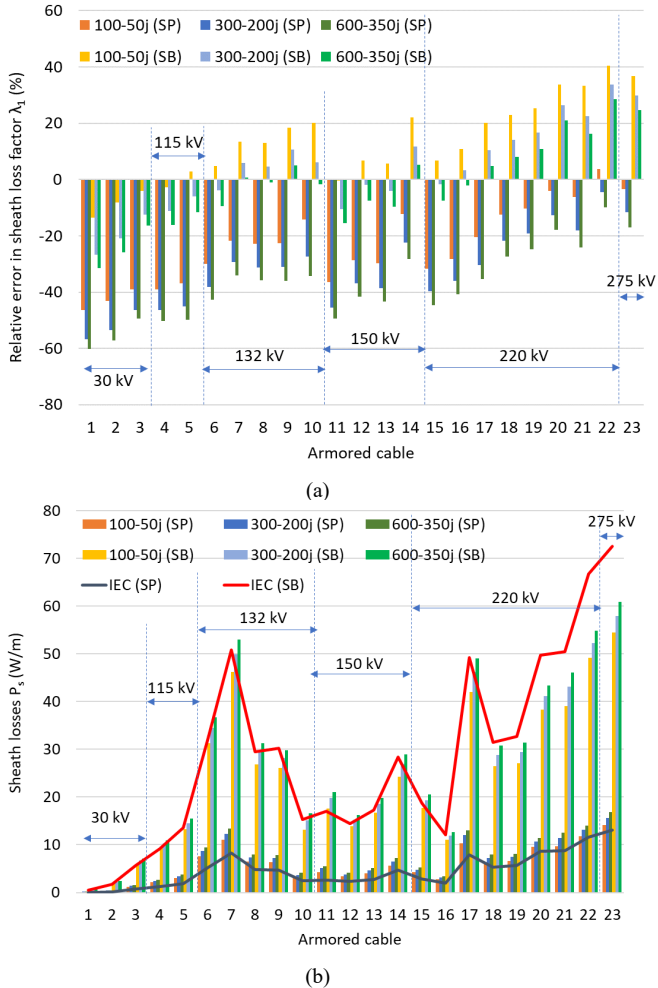

(a)

(b)

Fig. 8. Armored cables in SP and SB (Cu conductor): (a) relative error in $\lambda_1$ between FEM and IEC results and (b) results for $P_s$ derived by IEC and FEM for different values of $\mu_r$ ($\theta_{amb}$ = 20 ºC).

On the other hand, opposite results are observed when sheaths are in SB and $\lambda''_1$ is not neglected, since now the IEC formulae usually overestimate $\lambda_1$, as observed in Fig. 8a and 8b, where copper conductors and different values of $\mu_r$ for the armor wires are considered (results labeled as SB). In this case, it is again observed that the values of $P_s$ derived from FEM increase with $\mu_r$ (Fig. 8b), being between 1.2 and 2 times greater than in unarmored ones. This is partially explained due to an increase of circulating currents when the armor is considered, since the current induced in the sheaths is about 1.1 to 1.4 times greater than in unarmored cables, depending on the value of $\mu_r$, as observed in Fig. 9 where the values of $I_c$ and $I_s$ derived from FEM simulations are represented.

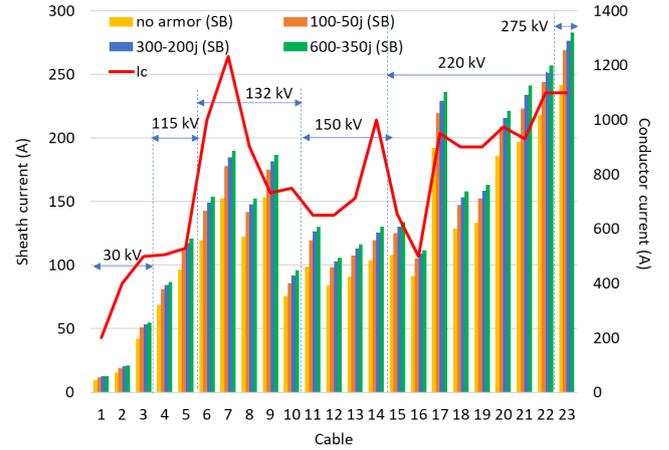

Fig. 9. Sheath and conductor current in unarmored and armored cables in SB (Cu conductor) derived by FEM for different values of $\mu_r$ ($\theta_{amb}$ = 20 ºC).

Finally, Fig. 10 shows the relative error of [2] over FEM in the estimation of $P_c$ for armored cables with copper conductors (similar results for aluminum). It is clearly observed how [2] underestimates $P_c$ when sheaths are both in SB and SP configurations. In fact, FEM results show that $P_c$ increases with $\mu_r$, so that it can be up to 10 % higher than in unarmored cables (depending on the cable geometry and $\mu_r$).

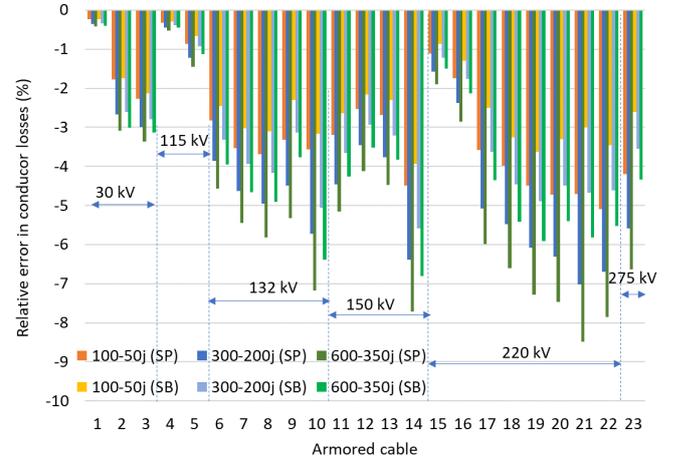

Fig. 10. Armored cables in SP and SB (Cu conductor): relative error in $P_c$ between FEM and IEC results for different values of $\mu_r$ ($\theta_{amb}$ = 20 ºC).

From this analysis important conclusions are to be remarked since they strongly influence the results derived from the EM. On one side, it is clear that (13) should be considered in the computation of $\lambda_1$ for both SP and SB configurations, although it does not properly collect the behavior of sheath eddy-current losses in three-core cables, either in armored or unarmored cables. The reason for this behavior is not only the presence of the armor, but also the twisting of the conductors. These circumstances modify the magnetic field inside the cable, increasing both circulating and eddy currents in the sheaths. Consequently, (2) and (13) need to be adjusted to include properly the effects of both aspects in the computation of $P_s$.

On the other hand, the higher values obtained in the conductor losses when the armor is considered may be interpreted as an increase in the conductor AC resistance, so

(11) needs to be also adjusted for improving the accuracy in the evaluation of $P_c$. Thus, the improvements in the computation of $P_s$ and $P_c$ will increase the accuracy of EM, as detailed next.

## V. Improvement of losses estimation in SB cables

Since the EM in [13] is based on a differential method, (11)-(13) should provide accurate values for $P_s$ and $P_c$. Otherwise, the numerical estimation of $P_a$ would contain part of the losses associated with the conductors and the sheaths, leading to an unacceptable assignment. The results obtained in the previous section show that these expressions need to be improved, so, for sheaths in SB (the preferred bonding configuration in offshore applications [6], [15], [16]), [1] proposed to make the following approaches for improving the accuracy of $P_a$ through the EM:

$$P_{c_1}^J \approx 1.02 \cdot P_{c_0}^J, \qquad (21)$$

$$P_{s_1}^{ec} \approx 1.35 \cdot P_{s_0}^{ec}, \qquad (22)$$

leading to the following expression for $P_a$ instead of (6):

$$P_a = \Delta P_m - 3R_c I_{c0}^2 (0.02 + 0.35 \cdot \lambda_1'') - 3R_s (I_{s1}^2 - I_{s0}^2). \quad (23)$$

This approach provided better results than the original EM in [13] for the 8 cables analyzed in [1]. However, they were not always so accurate when extended to more cables, especially for higher values of $\mu_r$. This is now further improved thanks to the conclusions of Section III together with the VL tool, since it is now possible to make an in-depth parametric analysis in numerous three-core armored cables. Through this tool, the set of 368 cases presented in Table II were deeply analyzed, providing valuable data for improving the knowledge of the phenomena involved in such complex cables, and finding the parameters that mostly influence $P_c$ and $P_s$ in either armored or unarmored cables. As a result, new analytical expressions are proposed to replace the numerical correction factors in (21)-(23), leading to an important improvement in the accuracy of the EM for the experimental estimation of the armor losses, as detailed next.

### A. Conductor Losses

In the previous section it was concluded that $R_c$ is affected by the presence of the armor. This may be assumed as an increase in the proximity effect in the conductor, so here we propose to apply a correction factor $f_c$ to the proximity effect factor $y_p$ in order to include this effect in $R_c$:

$$R'_c = R_c^{DC}(1 + y_s + y_p \cdot f_c), \qquad (24)$$

$$f_c = f\left(CP, \frac{s}{D_a}, \frac{d}{d_a}, \frac{d_a}{D_a}, \mu'\right). \qquad (25)$$

### B. Sheath Losses

Since it is not possible to evaluate separately circulating and eddy-current losses, a sheath equivalent resistance $R_s^{eq}$ is proposed:

$$P_s = 3 R_s^{eq} I_s^2. \qquad (26)$$

This resistance must be obtained for armored ($R_{s_1}^{eq}$) and unarmored ($R_{s_0}^{eq}$) cables for evaluating $P_{s_1}$ and $P_{s_0}$ in the EM. Thus, $R_{s_0}^{eq}$ is obtained from $R_s^{DC}$ by applying a correction factor $y_c$ in the form of

$$R_{s_0}^{eq} = R_s^{DC}(1 + y_c), \qquad (27)$$

$$y_c = f\left(LF, \frac{d_c}{d}, \frac{2s}{d}\right), \qquad (28)$$

being $d_c$ the conductor diameter and $LF$ the lay factor [24], defined as

$$LF = \sqrt{1 + \left(\frac{2\pi c}{L_c}\right)^2}. \qquad (29)$$

On the other hand, the effect of the armor is included with a second correction factor $y_a$, so that

$$R_{s_1}^{eq} = R_{s_0}^{eq}(1 + y_a), \qquad (30)$$

$$y_a = f(CP, \mu'). \qquad (31)$$

### C. Analytical Correction Factors for $P_c$ and $P_s$

The parameters in (25), (28) and (31) have been selected after an in-depth parametric analysis developed in the 368 cases presented earlier. Once done, different analytical expressions are obtained for $f_c$, $y_c$, and $y_a$ by means of a demo version of software Eureqa [27]. This software is a symbolic regression program that provides approximate expressions to fit a set of numerical results by means of a genetic algorithm. In this case, the inputs for this software are the values of the parameters in (25), (28) and (31), as well as the values of $P_c$ and $P_s$ obtained by FEM simulations for all the cables analyzed. As a result, it provides numerous expressions for each factor. These expressions are ordered depending on the fitness function selected. Here, we propose the most compact expressions obtained from the analysis of a subset of 184 cases that include armored and unarmored cables (only in SB):

$$f_c = 1 + \left(\frac{100}{1.5 + CP}\frac{s}{D_a}\frac{d}{d_a} - 60\right)\left(\frac{d_a}{D_a}\right)^2 \ln \mu', \qquad (32)$$

$$y_c = \frac{6.6 + 2.6\sqrt{LF^2 - 1}}{5.2\frac{d_c}{d} + \left(\frac{2s}{d}\right)^4}, \qquad (33)$$

$$y_a = 6 \cdot 10^{-3}(\ln \mu' - CP^2). \qquad (34)$$

The accuracy achieved with these expressions are depicted in Fig. 11 and Fig. 12 for $\theta_{amb} = 20$ ºC. In particular, Fig. 11 represents the error, relative to FEM simulations, in the estimation of $P_c$ obtained when using the corrected value of $R_c$ provided by (24) and (32). As can be seen, a great improvement has been achieved compared to those previously observed in Fig. 10, since differences are now below 2 %, either in armored or unarmored cables, independently of the material employed in the conductors or the permeability of the armor wires.

Another improvement is obtained regarding sheath losses in armored and unarmored cables by means of (26), (30), (33) and (34), as seen in Fig.12. The relative differences decrease from the 20–40 % observed in Fig. 8a to 1.5 % at most in Fig. 12.

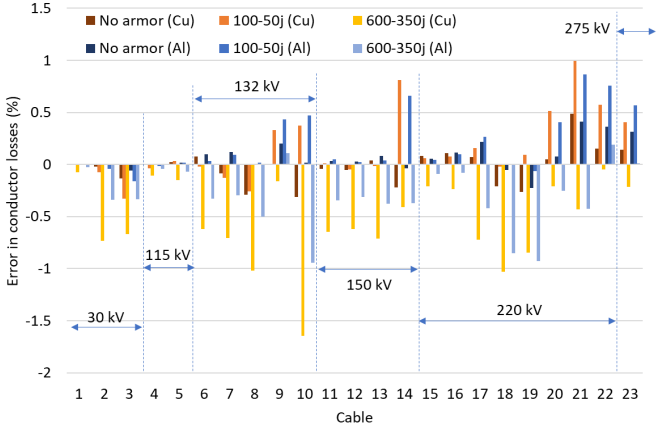

Fig. 11. Relative error in the estimation of $P_c$ for armored and unarmored cables in SB (Cu and Al conductors) ($\theta_{amb}$ = 20 ºC).

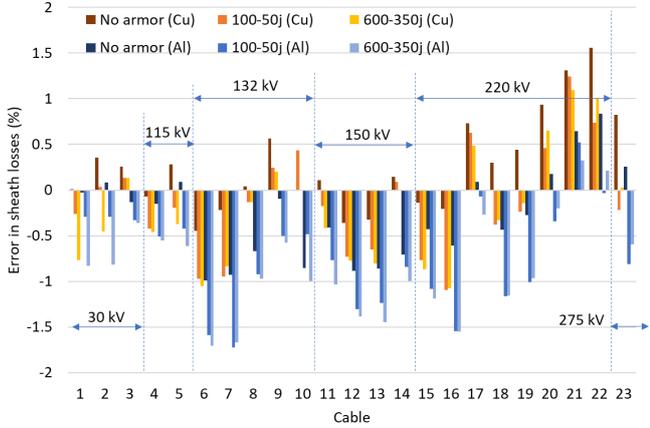

Fig. 12. Relative error in the estimation of $P_s$ for armored and unarmored cables in SB (Cu and Al conductors) ($\theta_{amb}$ = 20 ºC).

### D. Improved EM

Thanks to these new approaches for estimating $P_c$ and $P_s$, the EM presented in [13] can be greatly improved by just replacing (8)-(10) by

$$\Delta P_c^J = 3 R_c^{DC} y_p I_c^2 (f_c - 1), \quad (35)$$

$$\Delta P_s^J = 3 R_s^{DC}(1 + y_c)[(1 + y_a) I_{s_1}^2 - I_{s_0}^2], \quad (36)$$

$$\Delta P_s^{ec} = 0. \quad (37)$$

where all the elements of the power cable are assumed to be at ambient temperature during all tests.

The main results when applying these new expressions in the estimation of $P_a$ for the 184 cable cases derived from the 23 geometries presented earlier are shown in Fig. 13 for a temperature of 20 ºC, where it is also included the results obtained by the original EM [13] and the previous approach presented in [1]. As can be seen, although [1] provides better results than [13] in most of the cases, the error relative to 3D-FEM simulations is high for the smallest cables. However, the new approach (vertical bars) shows excellent results since the maximum error is always within ±10 %.

To extend the validation of this new proposal, a new set of 80 cables geometries (ranging from 35 mm², 10 kV to 2000 mm² to 275 kV) were added to the 23 presented earlier. Some of their main geometrical parameters were obtained from manufacturer brochures and others had to be completed with [6] and well-known examples available in the literature. This set of 103 cables were later extended by considering copper and aluminum for the conductors. This was performed by randomly varying (within a certain range) some of their geometrical parameters ($L_c$, $L_a$, $N$, $d_a$, etc.), $\mu'$ and $\mu''$, as well as the ambient temperature ($\theta_{amb}$) to include its influence on $R_c^{DC}$, $R_s^{DC}$ and $R_a^{DC}$, obtaining up to 645 cable cases. The results obtained in this set by applying [1], [13] and the new proposal are represented in Fig. 14 (cable labels ordered in increasing voltages and cross sections). Table III also show relevant results for some of the analyzed cases when tests are performed at different values of $\theta_{amb}$.

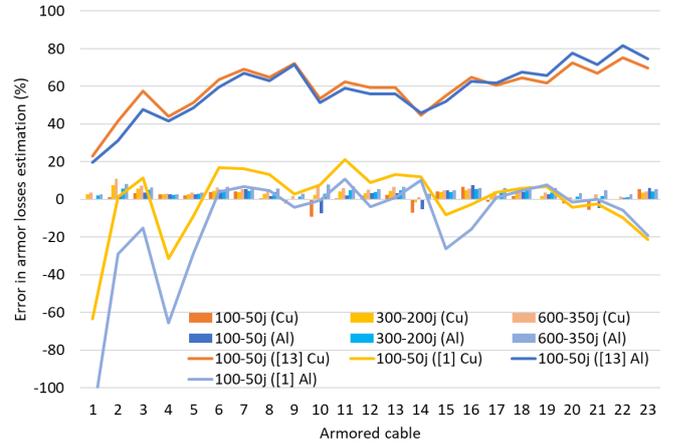

Fig. 13. Relative error in the estimation of $P_a$ for armored cables in SB (Cu and Al conductors) with [1], [13] and the new approach ($\theta_{amb}$ = 20 ºC).

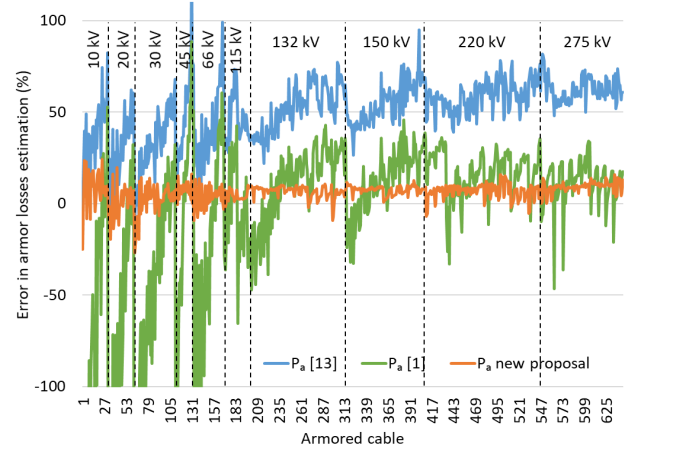

Fig. 14. Relative error in the experimental estimation of $P_a$ for 645 armored cables in SB with [1], [13] and the new approach.

TABLE III. RESULTS IN THE ESTIMATION OF $P_a$ FOR SOME OF THE CABLES AT DIFFERENT TEMPERATURES (CU CONDUCTOR AND $\mu_r$ = 300 – $j$200)

| Volt. (kV) | $I_{s0}$ (A) | $I_{s1}$ (A) | $\theta_{amb}$ (ºC) | $R_c^{DC}$ | $R_s^{DC}$ | $\Delta P_m$ | $\Delta P_c^J$ | $\Delta P_s^J$ | $P_a$ | Err. (%) |
|---|---|---|---|---|---|---|---|---|---|---|
| | | | | (Ω/km) | | (W/m) | | | | |
| 30 | 8.97 | 12.09 | 30 | 0.128 | 1.194 | 0.868 | 0.042 | 0.308 | 0.518 | 2.3 |
| 132 | 122.5 | 148.0 | 20 | 0.0186 | 0.345 | 21.51 | 2.1 | 9.76 | 9.66 | 2.1 |
| 150 | 95.26 | 122.4 | 30 | 0.0251 | 0.329 | 15.06 | 1.20 | 7.933 | 5.93 | 3.7 |
| 275 | 258.4 | 293.7 | 2 | 0.0069 | 0.180 | 27.13 | 1.915 | 14.16 | 11.06 | 1.0 |
| | $I_s$ | $\theta_c$ | $\theta_s$ | $\theta_a$ | | $\Delta P_m$ | $\Delta P_c^J$ | $\Delta P_s^J$ | $P_a$ | Err. |
| 150 | Unarm. 88.8 | 55.2 | 49.0 | 40.8 | | 15.64 | 1.99 | 7.49 | 6.16 | 5.6 |
| | Armor. 112.9 | 62.8 | 56.2 | 47.8 | | | | | | |
| 275 | Unarm. 204.4 | 77.9 | 70.1 | 56.5 | | 28.86 | 3.34 | 12.62 | 12.89 | 6.6 |
| | Armor. 230.6 | 90.2 | 81.9 | 67.7 | | | | | | |

It can be seen how [1] provides better results than the EM only for cables above 115 kV. However, the new proposal shows excellent results in all the cases, especially for the smallest cables (from 10 kV to 30 kV) when compared to those derived by [1] and [13], since its maximum error is always below ±20 %. Furthermore, for cables above 45 kV the relative error is even lower (±10 % at most). Differences slightly increase, but remain below ±10 %, when the cable achieves a different thermal regimen during both tests, as shown in Table III for a 150 kV and a 275 kV cable, where $\theta_c$, $\theta_s$ and $\theta_a$ are the conductor, sheath and armor temperature, respectively. All these results show how the analytical expressions here proposed substantially improve the experimental evaluation of $P_a$ by [13], as well as the estimation of $P_c$ and $P_s$ by [2], being applicable to the great majority of the three-core armored cables.

## VI. Conclusions

The proper allocation of losses (conductor, sheaths and armor) in three-core lead-sheathed armored power cables is a challenging task. This paper proposes an improvement in the estimation of the conductor and sheath losses with the addition of new analytical expressions that compensate the values obtained by the IEC standard [2]. This way, the determination of the armor losses through the experimental method [13] is greatly improved. These expressions have been obtained by analyzing a set of 368 three-core armored cable configurations through an ad hoc 3D FEM-based application interface (named Virtual Lab) that performs a virtualization of the experimental tests and cable variants (armored/unarmored, single point or solid bonding). Its accuracy has been tested in an additional set of 645 cable configurations, leading to a remarkable improvement in the estimation of the sheath and conductor losses for any size and voltage of the cable: differences over 3D-FEM less than 2 % for the conductor losses and less than 1.5 % for sheath losses. Moreover, a substantial improvement is also achieved in the experimental determination of the armor losses when applying the proposed expressions, with differences over 3D-FEM less than 20 % in cables below 45 kV and 10 % for cables above 45 kV.

With these new expressions the way to improve the IEC standard 60287 in an effective way is paved, leaving for further research the development of direct analytical expressions for the armor losses $P_a$ and its loss factor $\lambda_2$.